\documentclass[supplement]{ptptex}
\usepackage{graphicx}

\title{
Non-equilibrium Critical Dynamics and Precursory Phenomena \\
in Color Superconductivity 
}

\author{
Masakiyo \textsc{Kitazawa}$^a$,
Tomoi \textsc{Koide}$^b$,
Teiji \textsc{Kunihiro}$^c$ and
Yukio \textsc{Nemoto}$^d$
}

\inst{
$^a$ Deparment of Physics, Kyoto University, Kyoto, 606-8502, Japan,\\
$^b$ Institute f\"ur Theoretische Physik, J.W.Goethe Universit\"at, D-60054
Frankfurt, Germany,\\
$^c$ Yukawa Institute for Theoretical Physics, Kyoto University, Kyoto,
606-8502, Japan,\\
$^d$ RIKEN BNL Research Center, BNL, Upton, NY 11973
}

\abst{
We derive an effective equation
to describe the non-equilibrium dynamics of the pair field 
of color superconductivity near the critical temperature.
We also discuss how critical dynamics affects the specific heat
and density of state above the critical temperature.
}

\newcommand{\bfk}{\mbox{{\boldmath $k$}}}
\newcommand{\bfp}{\mbox{{\boldmath $p$}}}

\begin{document}

\maketitle

The strong coupling nature of QCD at low energy can enhance
the strength of the fluctuation of the pair field 
in the color superconductivity (CS) at low density
around the critical temperature $T_c$.
In fact, it was shown 
that there appear large pair fluctuations in rather 
wide range of temperatures near $T_c$\cite{ref:KKKN1}.
In this talk, we first derive an effective equation 
to describe the non-equilibrium critical dynamicis 
of the pair field\cite{ref:KK}.
Then, we consider effects of the pair fluctuation on some observables
and show that anomalous behaviors are clearly seen 
in these observables even well above $T_c$\cite{ref:KKKN2,ref:KK}.

\begin{figure}[t]
\begin{center}
\includegraphics[width=6.5cm]{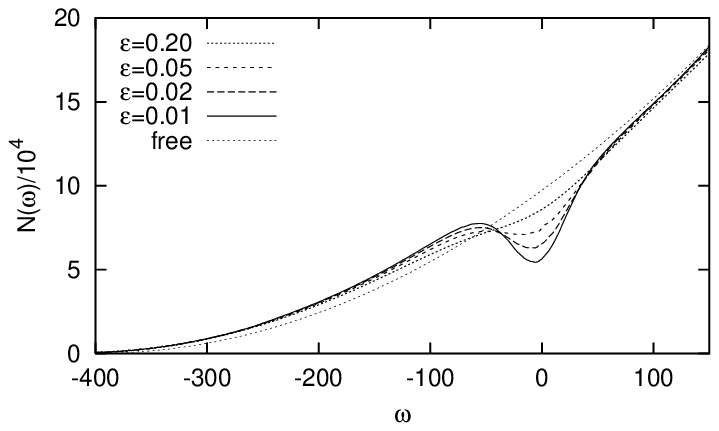}
\includegraphics[width=6.5cm]{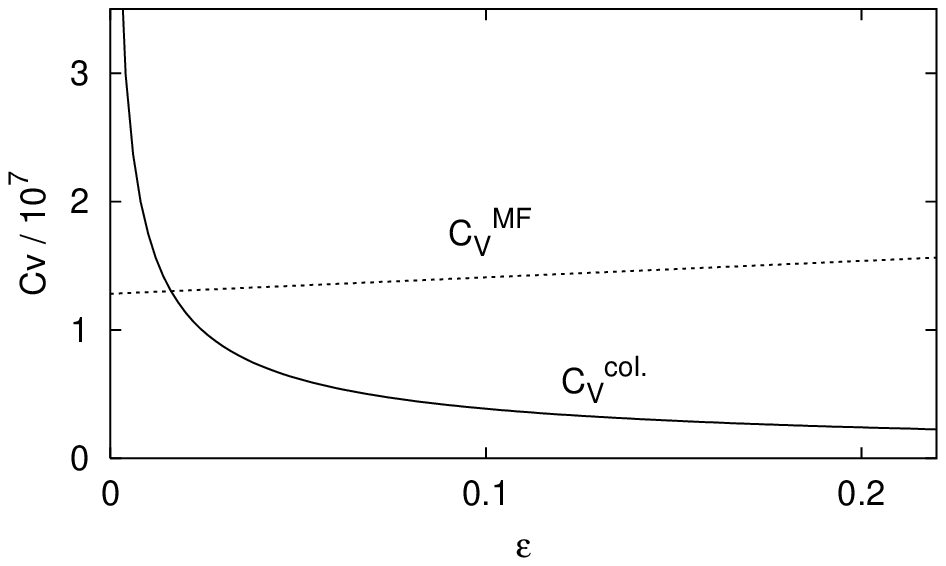}
\caption{
{\bf Left panel:} 
The density of state above $T_c$ for various reduced temperature
$\epsilon \equiv (T-T_c)/T_c$ with $\mu=400$MeV\cite{ref:KKKN2}.
{\bf Right panel:} 
The specific heat $C_{\rm v}$ with $\mu=400$MeV\cite{ref:KK}.
The contribution from the collective mode $C_{\rm v}^{\rm col.}$ 
is enhanced anomalously from $\epsilon = 0.05 \sim 0.1$.
}
\label{fig}
\end{center}
\end{figure}

Since we are interested in relatively low energy regions,
we shall employ the Nambu-Jona-Lasinio model with two-flavors
as an low energy effective theory of QCD.
In the random phase approximation,
the response function of the pair field reads
$
D^R( \bfk,\omega ) = 
-Q( \bfk,\omega ) / \left( G_C^{-1} + Q( \bfk,\omega ) \right),
$
with $Q(\bfk,\omega)$ denoting the lowest 
polarization function\cite{ref:KKKN1}.
There appears anomalous behavior in $D^R$ around $T_c$
in accordance with the growth of the fluctuation
of the diquark pair field.
The large fluctuations in turn contribute to various observables.
For example, effects on the quark self-energy and density of state
near $T_c$ is explored in \citen{ref:KKKN2} and shown that 
the {\it pseudogap} is formed above $T_c$ 
as a precursory phenomenon of the CS
(see, the left panel of Fig.~\ref{fig}).
It will be also shown later that the pair fluctuation can affect
the specific heat\cite{ref:KK}.

Before the calculation of observables, we first consider 
the non-equilibrium dynamics of the pair field near $T_c$.
The dynamics of the pair field is well described by
the pole of the collective mode (the {\it soft mode} of the CS)
$\omega=\omega(\bfk)$.
Since the response function $D^R$ diverges at $\omega=\omega(\bfk)$,
the collective mode is given by
$ \Xi^{-1} \equiv G_C^{-1} + Q( \bfk,\omega(\bfk) ) = 0 $.
Here, we expand $\Xi^{-1}$ around $\omega=|\bfk|=0$ and $T=T_c$:
\begin{eqnarray}
\Xi^{-1}( \bfk,\omega ) \equiv G_C^{-1} + Q( \bfk,\omega ) 
\simeq A\epsilon + B |\bfk|^2 + C \omega,
\label{eqn:Xi_expansion}
\end{eqnarray}
where $\epsilon \equiv (T-T_c)/T_c$ is the reduced temperature
and
$ A = T\partial Q({\bf 0},0) / \partial T |_{T=T_c} $,
$ B = \partial Q({\bf 0},0) / \partial \bfk^2 |_{T=T_c} $ and
$ C = \partial Q({\bf 0},0) / \partial \omega |_{T=T_c} $.
Then, the pole of the soft mode $\Xi^{-1}=0$ reads
$\omega = -(A/C)\epsilon -(B/C)|\bfk|^2$.
A numerical calculation shows that this simple equation
well reproduces the position of pole 
in $T \lesssim 1.3T_c$ and $k \lesssim 150$MeV\cite{ref:KK}.

In the linear response theory,
the dynamics of the pair field $\Delta(\bfk,\omega)$ 
is given by $\Xi(\bfk,\omega)\Delta(\bfk,\omega)=0$\cite{ref:KKKN1};
which, with expansion (\ref{eqn:Xi_expansion})
corresponds to the linearized time-dependent Ginzburg-Landau equation
of $\Delta$\cite{ref:KK}.
The time evolution of $\Delta$ is easily calculated
with this effective equation.
It is worth mentioning that the coefficient $C$ 
is not pure imaginary which means that
the collective mode is a damped oscillating mode in this case.
This is different from the soft mode in the weak coupling limit 
which is found to be an overdamped mode.
One can show that the particle-hole asymmetry 
gives rise to the finite real part of $C$\cite{ref:KK}.

Next, we turn to the discussion on the specific heat $C_{\rm v}$,
which is given by the themodynamic potential $\Omega$;
$ C_{\rm v} = -T \partial^2 \Omega / \partial T^2$.
The contribution to $\Omega$ from the collective modes $\Omega_{\rm col.}$
may be given by the summation of the connected ring diagrams,
\begin{eqnarray}
\Omega_{\rm col.}
= 
3 T\sum_n \int \frac{ d^3\bfp }{ (2\pi)^3 } 
\log (G_C^{-1} + Q( \bfk,i\nu_n ) ) 
=
3 T\sum_n \int \frac{ d^3\bfp }{ (2\pi)^3 } 
\log \Xi^{-1}( \bfk,i\nu_n ).
\nonumber
\end{eqnarray}
Then, 
$C_{\rm v}^{\rm col.} = -T \partial^2 \Omega_{\rm col.} / \partial T^2$
is the contribution to the specific heat from the collective modes.
Note that the calculation of $C_{\rm v}^{\rm col.}$ is well
simplified using the expansion (\ref{eqn:Xi_expansion}).
Temperature dependece of $C_{\rm v}^{\rm col.}$ is shown 
in the right panel of Fig.~\ref{fig} together with the specific heat 
of the free quark system $C_{\rm v}^{MF}$;
the sum of them $ C_{\rm v} = C_{\rm v}^{MF} + C_{\rm v}^{\rm col.} $
gives a total specific heat of the system.
From the figure, one sees that $C_{\rm v}^{\rm col.}$ is enhanced
anomalously as above $T_c$ as $\epsilon = 0.05 \sim 0.1 $. 
The enhancement of the specific heat in such a wide range of $T$
might affect the cooling process of newly borned compact stars.
It is also interesting to investigate other precursory phenomena
in various observables.


%


\begin{thebibliography}{99}
\bibitem{ref:KKKN1}
M.~Kitazawa, T.~Koide, T.~Kunihiro and Y.~Nemoto,
Phys.\ Rev.\ D {\bf 65}, 091504(R) (2002).
%
\bibitem{ref:KK}
M.~Kitazawa and T.~Kunihiro,
in preparation.
%
\bibitem{ref:KKKN2}
M.~Kitazawa, T.~Koide, T.~Kunihiro and Y.~Nemoto,
hep-ph/0309026.
%


\end{thebibliography}
\end{document}